# Injection Bias Reduction Techniques in Quantitative Angiography Using Patient-Specific Phantoms of Intracranial Aneurysm

Parmita Mondal[1,2], Kyle A Williams[1,2], Parisa Naghdi[1,2], Ahmad Rahmatpour[1,2], Mohammad Mahdi Shiraz Bhurwani[3], Swetadri Vasan Setlur Nagesh[2], Ciprian N Ionita[1,2,3]
[1]Department of Biomedical Engineering, University at Buffalo, Buffalo, NY 14260
[2]Canon Stroke and Vascular Research Center, Buffalo, NY 14203
[3]QAS.AI Inc, Buffalo, NY 14203

## Abstract

In intracranial aneurysm (IA) treatment, digital subtraction angiography (DSA) monitors device-induced hemodynamic changes. Quantitative angiography (QA) provides more precise assessments but is limited by hand-injection variability. This study evaluates correction methods using in vitro phantoms that mimic diverse aneurysm morphologies and locations, addressing the 2D and temporal limitations of DSA.

We used a patient-specific phantom to replicate three distinct IA morphologies at various Circle of Willis points: the middle cerebral artery (MCA), anterior communicating artery (ACA), and the internal carotid artery (ICA), each varying in size and shape. The diameters of the IA at MCA, ACA and ICA are 10.1, 10 and 7 millimeters, respectively. QA parameters for both non-stenosed and stenosed conditions were measured with 5ml and 10ml boluses over various injection durations to generate time density curves (TDCs). To address the variability in injection, several singular value decomposition (SVD) variants—standard SVD (sSVD) with Tikhonov regularization, block-circulant SVD (bSVD), and oscillation index SVD (oSVD)—were applied. These methods enabled the extraction of IA impulse response function (IRF), peak height ($PH_{IRF}$), area under the curve ($AUC_{IRF}$), and mean transit time (MTT). We evaluated the robustness of bias-reducing methods by observing the invariance of these parameters with respect to the injection conditions, and the location and size of the aneurysm.

The application of SVD variants—sSVD, bSVD, and oSVD—significantly reduced QA parameter variability due to injection techniques. MTT plots demonstrated a consistent decrease in variability across various aneurysm configurations and injection durations. This reduction in bias and enhanced parameter invariance was evident across different aneurysm sizes and locations, indicating the robustness of SVD methods in standardizing neurovascular diagnostic measures.

SVD-based deconvolution improves neurovascular diagnostic accuracy by reducing bias and maintaining consistent results across varying aneurysm sizes and locations.

## Introduction

An IA is ballooning arising from a weakened area in the wall of a blood vessel in the brain, which if ruptured can cause subarachnoid hemorrhage (SAH), leading to severe outcomes [1]. The overall mortality due to aneurysmal SAH is 0.4 to 0.6% of all-cause deaths, with an approximate 20% mortality and an additional 30 to 40% morbidity in patients with known rupture. It is well known that hemodynamics plays a critical role in aneurysm rupture and healing. Thus, altering the hemodynamics is central to aneurysmal treatment management.

Interventionists employ endovascular methods to alter the hemodynamic conditions near the aneurysm treatment site, but the aneurysms' remote location and tortuosity often pose challenge for this intervention[2]. This results on the reliance on QA to have a quantitative assessment of the blood flow within aneurysm, via Digital Subtraction Angiography (DSA)[3-5]. However, one of the most significant limitations of QA is hand injection variability, which can notably impact the precision of QA estimations[6, 7]. It is caused due to different hand or mechanical injector methods. The deconvolution methods based on SVD can be used to deconvolve the arterial input function (AIF) from aneurysm TDCs, to help mitigate the effects introduced by injection variability.

We implemented deconvolution on the in-vitro acquired angiograms, obtained using the experimental setup described below. In this study, we use SVD based deconvolution methods to reduce the impact of hand injection variability on QA parameters. SVD is a mathematical technique used in linear algebra, which induces factorization of a matrix into three other matrices, which when multiplied together, reconstruct the original matrix[8, 9]. The different SVD variants used in this study are sSVD with Tikhonov regularization, bSVD and oSVD[10, 11]. In the sSVD approach, the AIF – the inlet TDC, is transformed into a Toeplitz matrix, whereas in bSVD and oSVD, the AIF is transformed into block circulant matrix. The transformed AIF matrix, undergoes SVD, generating IRF.

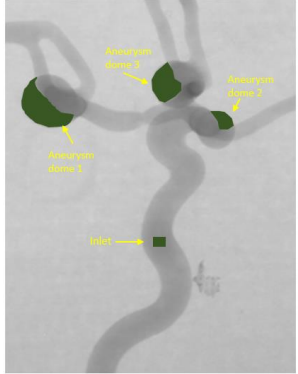

**Figure 1.** Aneurysm geometry is shown., where yellow arrows show the aneurysm dome and inlet ROI. Three aneurysm geometries are chosen for this study.

Figure 1 shows the three IA geometries at MCA, ACA and ICA, respectively. We repeat the above steps for each of all the deconvolution methods. The QA parameters evaluated from IRF are $PH_{IRF}$, $AUC_{IRF}$ and MTT. $PH_{IRF}$ which is calculated by taking the maximum value of IRF. $AUC_{IRF}$ which is calculated by integrating the IRF with respect to AIF. and MTT is calculated by dividing the $AUC_{IRF}$ by $PH_{IRF}$. Additionally, in oSVD we evaluate a metric called oscillation index (OI), which gives the variability in the IRF[12-14].

In this study, we use different SVD methods to deconvolve the AIF from aneurysm TDC, to generate IRF and evaluate the QA parameters, which would be devoid of injection variability. In this study, we use in-vitro acquired phantoms, to replicate real life scenarios where factors like noise, patient motion and X-ray induced artifacts are included[15]. We implement the method on both non-stenosed and stenosed parent arteries.

## Material and Methods

### 2.1 Experimental Setup

Our experimental setup was adapted from the methodology described by Ionita et al., using a patient-specific internal carotid artery phantom to replicate realistic physiological conditions. The phantom was connected to a cardiovascular waveform pump (Model 1407, Harvard Apparatus, MA), operating at 70 pulses per minute with a systole-to-diastole phase ratio of 30:70 and a stroke volume of 20 cc. A damper was integrated into the flow circuit to minimize the pulse wave effects generated by the pump. A contrast injector catheter and a pressure transducer were placed between the damper and the phantom to ensure accurate contrast injection and monitoring.

A 7Fr catheter connected to an automatic injector was used to deliver an iodine-based contrast agent. Fluid velocity at the main branch of the phantom was measured using an ultrasound probe paired with a transit-time flowmeter, yielding an average flow rate of 1.91 l/min and a carotid fluid velocity of 63.3 cm/s. To maintain experimental integrity, the used contrast was collected in a separate container to prevent contamination of the main reservoir, with strict control over flow back into the system.

Bolus injection volumes and durations were carefully regulated to evaluate their influence on imaging outcomes. A Canon Infinix bi-plane system captured DSA images at 15 frames per second. Boluses of 5 ml and 10 ml were injected at four rates—5 ml/s, 10 ml/s, 15 ml/s, and 20 ml/s—corresponding to injection durations of 1.0 s, 0.5 s, 0.33 s, and 0.25 s for the 5 ml bolus, and 2.0 s, 1.0 s, 0.66 s, and 0.5 s for the 10 ml bolus. Each condition was repeated three times to ensure data reliability.

To simulate different flow conditions, a 75% stenosis was introduced in the parent artery, and the same experimental parameters were applied to assess its effects. Figure 2 illustrates the study workflow.

### 2.2 Quantitative angiographic parameters analysis using deconvolution methods

We applied SVD-based deconvolution to in-vitro angiograms obtained from our experimental setup to minimize the effects of manual injection variability on QA parameters. The study utilized three SVD variants: sSVD with Tikhonov regularization, bSVD, and oSVD.

In the sSVD method, the arterial input function (AIF)—represented by the inlet TDC was converted into a Toeplitz matrix. For bSVD and oSVD, the AIF was transformed into a block circulant matrix. The transformed AIF matrices underwent singular value decomposition (SVD), breaking them down into three components: U (left singular vectors), V (right singular vectors), and S (a diagonal matrix containing non-negative singular values). This process was implemented using the Python SciPy linear algebra library. The impulse response function (IRF) of the aneurysm dome was then calculated and convolved with the AIF (Ca) to produce a reconvolved TDC (Qnew). This reconvolved TDC closely aligned with the original aneurysm TDC (Q).

This process was repeated for each deconvolution method. The QA parameters derived from the IRF included the peak height (PHIRF), area under the curve (AUCIRF), and mean transit time (MTT).

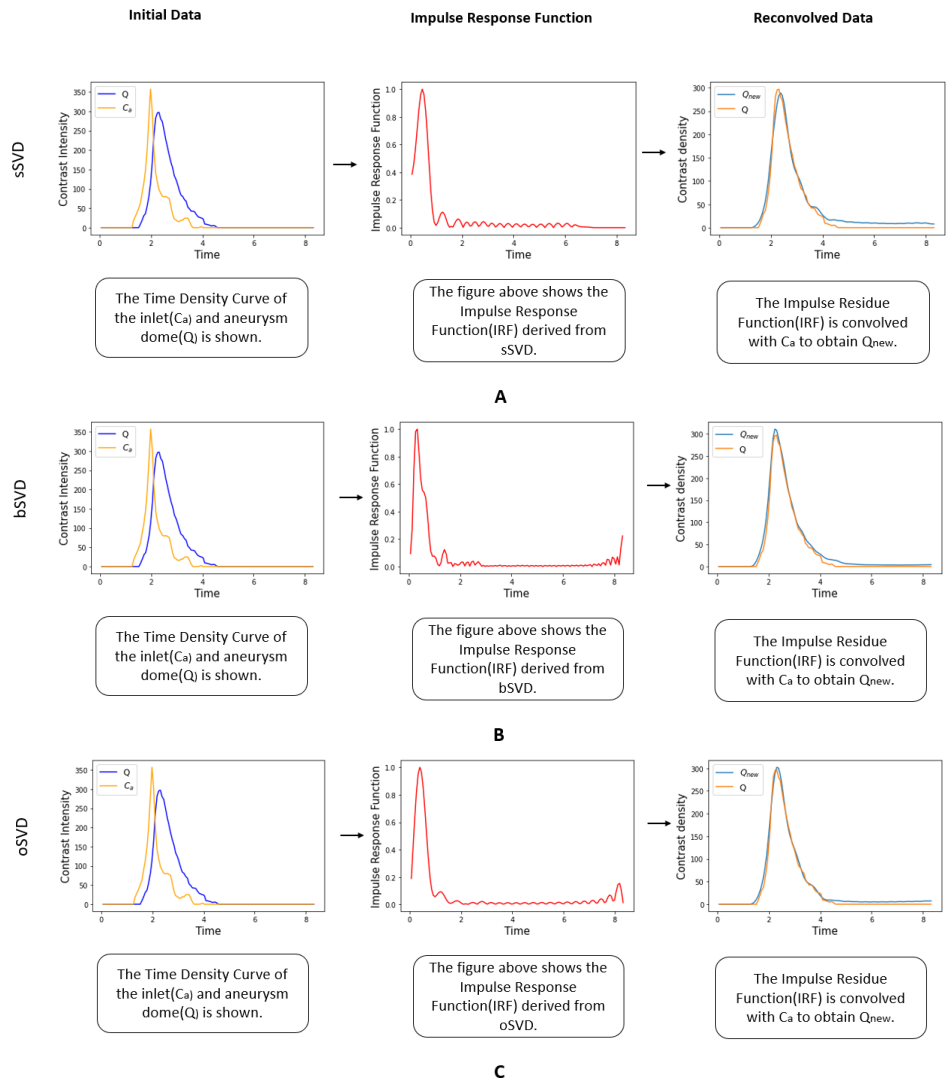

**Figure 2.** The implementation of sSVD (A), bSVD (B) and oSVD (C) is shown. We used in-vitro generated angiograms shown in Figure 1 to generate Time Density Curve (TDC) for the inlet and aneurysm dome, and implement sSVD, bSVD and oSVD. Q is the time density curve of the aneurysm dome; Ca is the time density curve of the inlet and Qnew is the time density curve obtained by convolving impulse residue function (IRF) with Ca. We evaluated IRF, to extract Quantitative Angiography (QA) parameters: PHIRF, AUCIRF and MTT. The leftmost column shows the original TDC of the aneurysm dome and inlet. The second column shows the IRF. The third column shows the generated Qnew, formed by convolving the inlet function with the IRF.

## Results

We have repeated the experiment with three different aneurysm sizes and locations, but we have provided the results for aneurysm geometry 1 (shown in Figure 1), having a 5ml bolus in Figure 3. Plot of MTT before and after the implementation of SVD variants are shown in Figure 3, demonstrating the reduction of slope after the implementation of SVD. The percentage reduction of slope for non-stenosed artery on implementation of sSVD is 92.56%, bSVD is 98.01% and oSVD is 98.45%. The percentage reduction of slope for stenosed artery on implementation of sSVD is 99.38%, bSVD is 74.25% and oSVD is 75.37%. Figure 4 shows the bar of the slope of MTT before and after the implementation of SVD variants for aneurysm geometry 1, having a bolus of 5ml and Table 1 shows $PH_{IRF}$, $AUC_{IRF}$ and $MTT_{SVD}$ for different parent artery (stenosed and non-stenosed) and injection duration, for the aneurysm geometry 1 having 5ml bolus using sSVD, bSVD and oSVD.

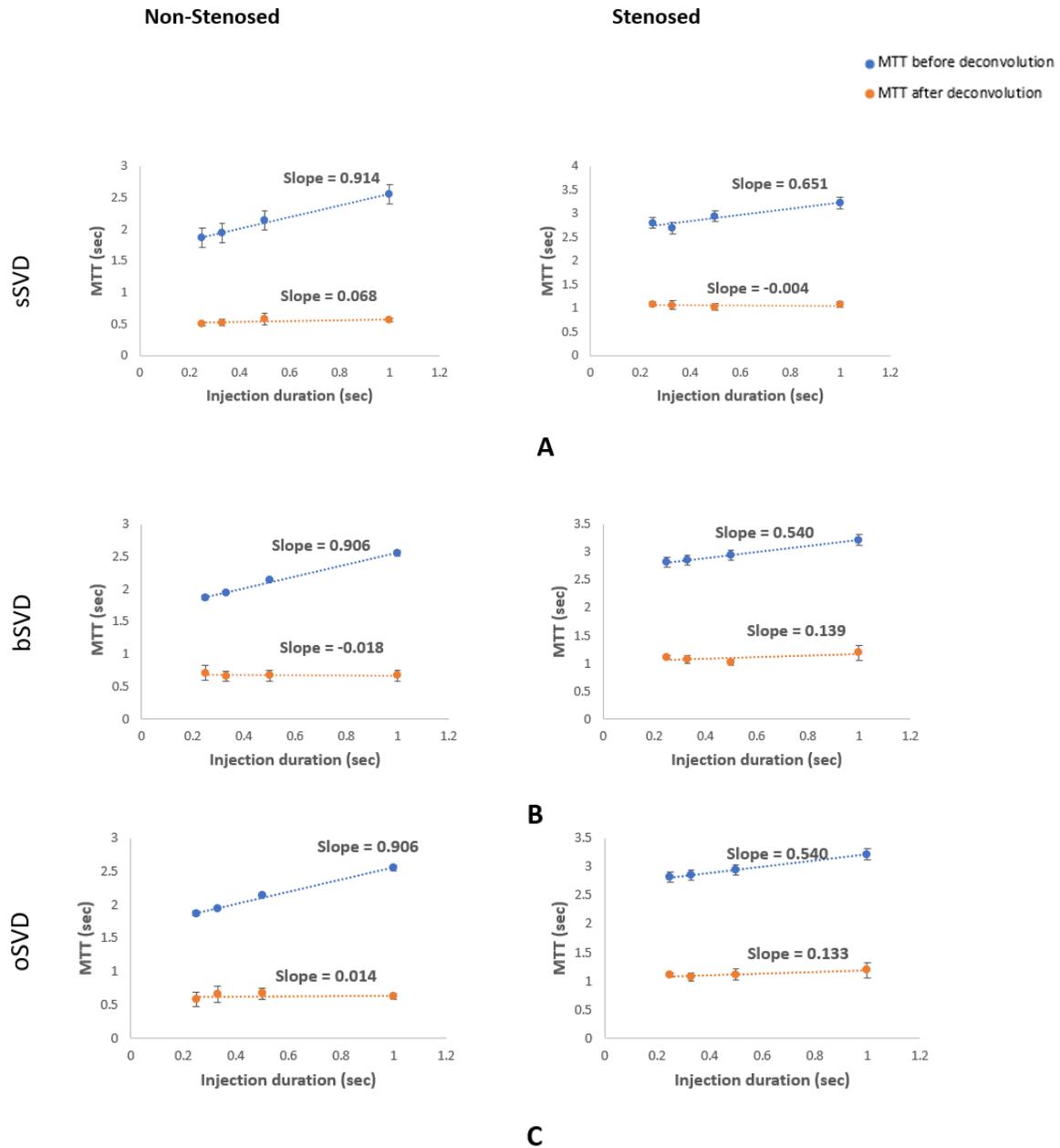

**Figure 3.** Plot of MTT after the implementation of sSVD (A), bSVD (B) and oSVD (C) is shown, at different parent artery (stenosed and non-stenosed) for aneurysm geometry 1 having 5ml bolus.

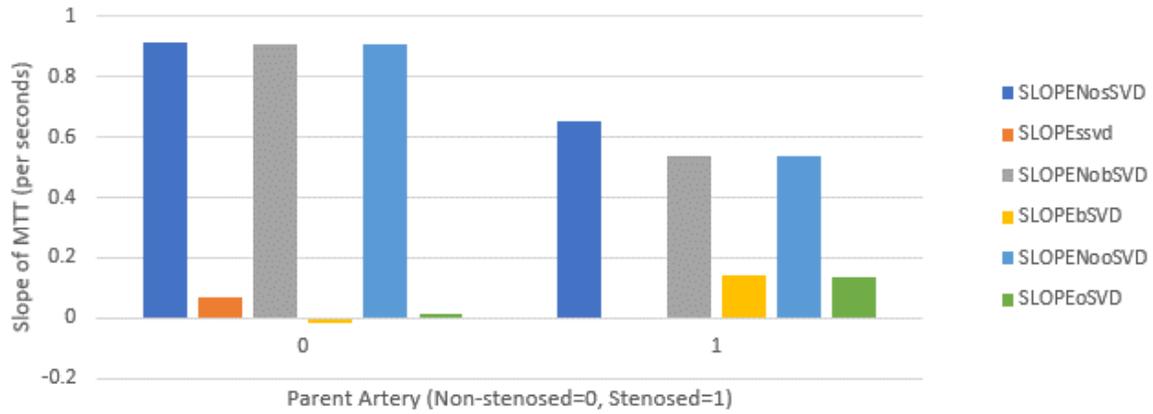

**Figure 4**. Bar plot showing the slope of MTT before and after the implementation of SVD variants: sSVD, bSVD and oSVD for non-stenosed(0) and stenosed(1) parent artery, for aneurysm geometry 1 having 5ml bolus. $SLOPE_{NosSVD}$ demonstrates the slope of MTT before implementation of sSVD. $SLOPE_{NobSVD}$ demonstrates the slope of MTT before implementation of bSVD. $SLOPE_{NooSVD}$ demonstrates the slope of MTT before implementation of oSVD. $SLOPE_{ssvd}$ gives the slope of MTT after implementation of sSVD. $SLOPE_{bsvd}$ gives the slope of MTT after implementation of bSVD and $SLOPE_{osvd}$ gives the slope of MTT after implementation of oSVD. The plot is generated from M1 model.

**Table 1:** Demonstrates $PH_{IRF}$, $AUC_{IRF}$ and $MTT_{SVD}$ for different parent artery and injection duration for aneurysm geometry 1 having 5ml bolus, using sSVD, bSVD and oSVD. $MTT_{SVD}$ is the mean transit time evaluated after the implementation of SVD and MTT is the mean transit time before the implementation of SVD. $SLOPE_{SVD}$ is the slope of $MTT_{SVD}$ and SLOPE is the slope of MTT. $DIFF_{SL}$ is the difference between $SLOPE_{SVD}$ and SLOPE. $AVG_{SL}$ is the average of the $DIFF_{SL}$ for each SVD variant. AVU is arbitrary volume unit.

| SVD variant | Parent Artery | Injection duration (sec) | $PH_{IRF}$ (AVU/min) | $AUC_{IRF}$ (AVU) | $MTT_{SVD}$ (sec) | MTT (sec) | $Slope_{SVD}$ | Slope | $DIFF_{SL}$ | $AVG_{SL}$ |
|---|---|---|---|---|---|---|---|---|---|---|
| sSVD | Non-Stenosed | 0.25 | 0.413 ± 0.154 | 0.204 ± 0.071 | 0.503 ± 0.024 | 1.863 ± 0.035 | 0.068 | 0.914 | 0.846 | 0.750 |
| | | 0.33 | 0.344 ± 0.049 | 0.178 ± 0.009 | 0.524 ± 0.047 | 1.943 ± 0.008 | | | | |
| | | 0.5 | 0.321 ± 0.036 | 0.185 ± 0.008 | 0.585 ± 0.091 | 2.135 ± 0.006 | | | | |
| | | 1 | 0.222 ± 0.017 | 0.124 ± 0.002 | 0.563 ± 0.035 | 2.553 ± 0.041 | | | | |
| | Stenosed | 0.25 | 0.079 ± 0.004 | 0.086 ± 0.009 | 1.088 ± 0.056 | 2.795 ± 0.019 | -0.004 | 0.651 | 0.655 | |
| | | 0.33 | 0.076 ± 0.01 | 0.08 ± 0.007 | 1.068 ± 0.097 | 2.691 ± 0.032 | | | | |
| | | 0.5 | 0.078 ± 0.008 | 0.08 ± 0.006 | 1.024 ± 0.078 | 2.942 ± 0.018 | | | | |
| | | 1 | 0.084 ± 0.012 | 0.089 ± 0.008 | 1.074 ± 0.066 | 3.221 ± 0.002 | | | | |
| bSVD | Non-Stenosed | 0.25 | 0.167 ± 0.023 | 0.115 ± 0.005 | 0.705 ± 0.114 | 1.871 ± 0.035 | -0.018 | 0.906 | 0.924 | 0.662 |
| | | 0.33 | 0.302 ± 0.043 | 0.195 ± 0.011 | 0.654 ± 0.077 | 1.946 ± 0.008 | | | | |
| | | 0.5 | 0.301 ± 0.027 | 0.200 ± 0.009 | 0.672 ± 0.086 | 2.135 ± 0.006 | | | | |
| | | 1 | 0.198 ± 0.013 | 0.131 ± 0.007 | 0.670 ± 0.085 | 2.553 ± 0.041 | | | | |
| | Stenosed | 0.25 | 0.08 ± 0.004 | 0.089 ± 0.007 | 1.113 ± 0.035 | 2.822 ± 0.019 | 0.139 | 0.540 | 0.401 | |
| | | 0.33 | 0.082 ± 0.007 | 0.088 ± 0.004 | 1.075 ± 0.074 | 2.852 ± 0.032 | | | | |
| | | 0.5 | 0.082 ± 0.007 | 0.083 ± 0.006 | 1.023 ± 0.056 | 2.942 ± 0.018 | | | | |
| | | 1 | 0.08 ± 0.011 | 0.094 ± 0.007 | 1.193 ± 0.134 | 3.221 ± 0.002 | | | | |
| oSVD | Non-Stenosed | 0.25 | 0.023 ± 0.015 | 0.005 ± 0.003 | 0.114 ± 0.051 | 1.871 ± 0.035 | 0.014 | 0.906 | 0.892 | 0.649 |
| | | 0.33 | 0.043 ± 0.043 | 0.011 ± 0.011 | 0.077 ± 0.077 | 1.946 ± 0.008 | | | | |
| | | 0.5 | 0.027 ± 0.027 | 0.009 ± 0.009 | 0.086 ± 0.086 | 2.135 ± 0.006 | | | | |
| | | 1 | 0.013 ± 0.01 | 0.007 ± 0 | 0.085 ± 0.028 | 2.553 ± 0.041 | | | | |
| | Stenosed | 0.25 | 0.08 ± 0.004 | 0.089 ± 0.007 | 1.113 ± 0.035 | 2.822 ± 0.019 | 0.133 | 0.540 | 0.407 | |
| | | 0.33 | 0.082 ± 0.007 | 0.088 ± 0.004 | 1.075 ± 0.074 | 2.852 ± 0.032 | | | | |
| | | 0.5 | 0.076 ± 0.007 | 0.084 ± 0.007 | 1.116 ± 0.096 | 2.942 ± 0.018 | | | | |
| | | 1 | 0.08 ± 0.011 | 0.094 ± 0.007 | 1.193 ± 0.134 | 3.221 ± 0.002 | | | | |

## Conclusion

This study demonstrates the ability of deconvolution based on SVD to mitigate hand injection variability, validating its use as a normalization method for QA analysis.